\documentclass[a4paper,12pt]{spieman}  % use this instead for A4 paper
\usepackage{amsmath,amsfonts,amssymb}
\usepackage{aas_macros}
\usepackage{graphicx}
\usepackage{xspace}
\usepackage{setspace}
\usepackage{tocloft}
\usepackage{lineno}
\usepackage{paralist}
\usepackage[free-standing-units]{siunitx}
\usepackage{subfig}
\DeclareSIUnit\pixel{px}

\title{Commissioning and improvements of the instrumentation, and
  launch of the scientific exploitation of OARPAF, the Regional
  Astronomical Observatory of the Antola Park}

  \author[a]{Ricci~D.}
  \author[b, c]{Tosi~S.}
  \author[d]{Cabona~L.}
  \author[d]{Righi~C.}
  \author[e,*]{La~Camera~A.}

  \author[b]{Marini~A.}
  \author[b,c]{Domi~A.}
  \author[b]{Santostefano~M.}
  \author[f]{Balbi~E.}
  \author[b]{Nicolosi~F.}

  \author[e]{Ancona~M.}
  \author[e]{Boccacci~P.}
  \author[b,g]{Bracco~G.}
  \author[b,c]{Cardinale~R.}
  \author[e,**]{Dellacasa~A.}
  \author[d]{Landoni~M.}
  \author[b,c]{Pallavicini~M.}
  \author[b,c]{Petrolini~A.}
  \author[b,c]{Schiavi~C.}
  \author[h]{Zappatore~S.}
  \author[d]{Zerbi~F.~M.}

  \affil[a]{INAF-Osservatorio Astronomico di Padova, Vicolo
    dell’Osservatorio 5, 35122 Padova, Italy.}

  \affil[b]{Università degli Studi di Genova, DIFI Dipartimento di
    Fisica, Via Dodecaneso 33, 16146, Genova, Italy.}

  \affil[c]{INFN-Sezione di Genova, Via Dodecaneso 33, 16146 Genova,
    Italy.}

  \affil[d]{INAF-Osservatorio Astronomico di Brera, Via E. Bianchi 46,
    23807, Merate (LC), Italy.}

  \affil[e]{Università degli Studi di Genova, DIBRIS Dipartimento di
    Informatica, Bioingegneria, Robotica e Ingegneria dei Sistemi, Via
    all’Opera Pia 13, 16145, Genova, Italy.}

  \affil[f]{Università degli Studi di Genova, DISTAV Dipartimento di
    Scienze della Terra, dell’Ambiente e della Vita, Corso Europa 26,
    16132, Genova, Italy.}

  \affil[g]{CNR-IMEN, via Dodecaneso 33, 16146 Genova, Italy.}

  \affil[h]{Università di Genova, DITEN Dipartimento di Ingegneria
    delle Telecomunicazioni, Elettrica, Elettronica e Navale, Via
    all’Opera Pia 11A, 16145, Genova, Italy.}

  \affil[*]{Now at Teiga Srls, Viale Brigate Partigiane 16, 16129,
    Genova, Italy.}

  \affil[**]{Now at Drafinsub Srl, Via al Molo Giano, 16128, Genova,
    Italy.}

\begin{document}
\maketitle

\newcommand{\stl}{\textsc{stl}\xspace}
\newcommand{\atik}{\textsc{atik}\xspace}
\newcommand{\stx}{\textsc{stx}\xspace}
\newcommand{\lhires}{\textsc{lhires iii} \xspace}
\newcommand{\flechas}{\textsc{flechas}\xspace}
\newcommand{\oarpaf}{\textsc{oarpaf}\xspace}
\renewcommand{\thefootnote}{\alph{footnote}}

\begin{abstract}
  The \oarpaf telescope is an $80\centi\meter$-diameter optical
  telescope installed in the Antola Mount Regional Reserve, in
  Northern Italy. This work presents the results of the
  characterization of the site, as well as developments and
  interventions that have been implemented, with the goal of
  exploiting the facility for scientific and educational purposes.
  During the characterization of the site, an average background
  brightness of $22.40 \, m_{AB}$ ($B$ filter) -- $21.14 \, m_{AB}$
  ($I$) per arcsecond squared, and a $1.5$--$3.0\arcsecond$ seeing,
  have been measured. An estimate of the magnitude zero points for
  photometry is also reported.
  The material under commissioning includes 3 CCD detectors for which
  we provide the linearity range, gain, and dark current; a 31 orders
  échelle spectrograph with $R\sim 8500$--$15000$, and a dispersion of
  $n = 1.39\times 10^{-6}\pixel^{-1}\lambda + 1.45\times
  10^{-4}\nano\meter/\pixel$, where $\lambda$ is expressed in
  $\nano\meter$.
  % ,
  % allowing radial velocity measurements up to
  % $15.8 \kilo\meter / \second$, and a long slit spectrograph.
  %
  The scientific and outreach potential of the facility is proven in
  different science cases, such as exoplanetary transits and active
  galactic nuclei variability. The determination of time delays of
  gravitationally lensed quasars, the microlensing phenomenon and the
  tracking and the study of asteroids are also discussed as
  prospective science cases.
\end{abstract}

% Include a list of up to six keywords after the abstract

% Include email contact information for corresponding author
{\noindent \footnotesize Corresponding author: Davide Ricci,
  \linkable{davide.ricci@inaf.it}}

%\begin{spacing}{2}   % use double spacing for rest of manuscript

%%%%%%%%%%%%%%%%%%%%%%%%%%%%%%%%%%%%%%%%%%%%%%%%%%%%%%%%%%%%%%%%%%%%
\section{Introduction}
\label{sec:introduction}

The Regional Astronomical Observatory of the Antola Park (\oarpaf),
located at \ang{44;35;28.46}N, \ang{9;12;12.49}E, in the territory of
Comune di Fascia\footnote{\url{https://goo.gl/maps/upEY3}}, is a
facility situated in the Ligurian Apennines at an altitude of about
$1\,480$ m above sea level.
The altitude, together with the extremely small light pollution
distinctive of the area, has promoted the set up of an
outreach-driven $80\centi\meter$, $f/8$ Cassegrain-Nasmyth
alt-azimuthal optical telescope, inaugurated in 2011, which is one of
the few and one of the largest optical telescopes in Italy available
in a public facility~\cite{2012ASInC...7....7F}.
It also features a fully equipped 60 seats conference hall,
a planetarium, a library, and guest rooms for observers.

Following the growing interest in the facility by college,
  master, graduate students and young researchers, the University of
Genova (Italy) took charge of scientific operations under an agreement
with the Antola Natural Park, who manages the facility on behalf of
Comune di Fascia.  An interdepartmental center was created to group
researchers of the University of Genova interested in astronomy and in
the related instruments and technology.  The center, named
ORSA\footnote{\url{http://www.orsa.unige.net/index.php/en/about/}},
standing for Observations and Research in the Science of Astronomy,
groups together researchers from the Department of Physics (DIFI), the
Department of Mathematics (DIMA), the Department of Chemistry and
Industrial Chemistry (DCCI), the Department of Informatics,
Bioengineering, Robotics and System Engineering (DIBRIS), the
Department of Telecommunications, Electric, Electronic and Naval
Engineering (DITEN),  and the Department for the Earth,
  Environment and Life Sciences (DISTAV). This ``bottom-up''
  process, triggered by the proactive initiatives of students that
  engaged senior researchers and faculty members, also led to the
  establishment of new scientific collaborations with the National
Institute for Astrophysics (INAF) and the National Institute for
Nuclear Physics (INFN).

So far, the telescope was only operated locally, mainly for
  outreach events, and it was initially provided with a sbig
    stl 11000m camera (hereafter \stl) and its embedded filter wheel
  with standard Johnson-Cousins $UBVRI$ photometric filters.  This
  camera was used during the site characterization, commissioning, and
  photometric observations.
  By means of this setup, we obtained first scientific results in several
  fields\cite{2016NCimC..39..284R, 2017PASP..129f4401R}, probing the
  scientific potential of the facility.
  In addition to the \stl, a \flechas échelle, an optical fiber
  spectrograph with a \textsc{atik {\normalfont Monochrome} 11000m}
  (hereafter \atik) is also available at the observatory.

  Despite the encouraging photometric results, \oarpaf suffered
  from several problems.
  During the first years, the original dome showed water
  leakages, posing a serious risk of damaging the telescope: therefore,
  a new dome was installed in 2020 by the Gambato
  company\footnote{\url{https://www.gambato.it/}}.

In 2017 DIFI participated in the call by the Italian Ministry for
Education, University and Research (MIUR) named ``Departments of
Excellence'', with a project aimed at launching new research lines in
astrophysics, as well as an augmented educational offer in the
field. \oarpaf played an important role in the project both for its
scientific usage and for its adequacy to form new students in the
field of observational optical astronomy. The project was
funded\footnote{\url{https://www.anvur.it/attivita/dipartimenti/}} and
two main upgrades were planned for \oarpaf: an improvement of the
instrumentation and a full remotization. In particular, the
instrumentation has been improved by the purchase of a \lhires
spectrograph, a Class-1 CCD \textsc{sbig stx 16801} (hereafter \stx),
a \textsc{fw7-stx} filter wheel with standard, $50\milli\meter$
$UBVRI$ and $H\alpha$ filters, a \textsc{stx}-Guider, and the
  new \textsc{ao-x} module for tip-tilt correction.

%%%%%%%%%%%%%%%%%%%%%%%%%%%%%%%%%%%%%%%%%%%%%%%%%%%%%%%%%%%%%%%%%%%%

The paper is organized as follows: the telescope is described in
Sect.~\ref{sec:telescope}.
We describe the steps undertaken for the site characterization using
the \stl in Sect.~\ref{sec:site}.
Furthermore, we detail the existing and the new scientific
material (Sect.~\ref{sec:commissioning}) and how we plan to
reconfigure it in a new multi-purpose instrument, with the goal of
using it for scientific purposes.
Then in Sect.~\ref{sec:pipeline} we describe the data reduction
pipeline that we are currently setting up.
Consequently, we introduce the remote control strategy
(Sect.~\ref{sec:remote-control}) that will complete the upgrade
process.
This will allow us to introduce astrophysical projects achievable with
the facility (Sect.~\ref{sec:science}).
Finally, we address its relevance for educational and outreach events
(Sect.~\ref{sec:outreach}).
A summary of results, as long as conclusions, are shown in
Sect.~\ref{sec:conclusions}.

%%%%%%%%%%%%%%%%%%%%%%%%%%%%%%%%%%%%%%%%%%%%%%%%%%%%%%%%%%%%%%%%%%%%
\section{Telescope}
\label{sec:telescope}

% ----------------------------------------------------------------
\begin{figure}[t]
  \centering
  \includegraphics[width=0.7\columnwidth]{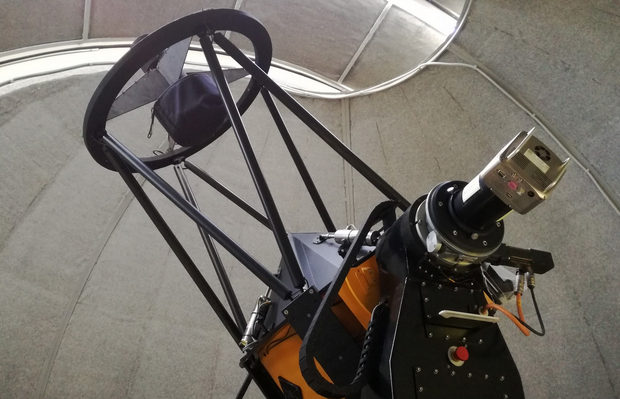}
  \caption{The \oarpaf telescope with the \stl on the derotated
    Nasmyth focus. }
  \label{fig:tel}
\end{figure}
% ----------------------------------------------------------------

The \oarpaf telescope (Fig.~\ref{fig:tel}) is a $0.8\meter$
alt-azimuthal Cassegrain-Nasmyth T0800-01 telescope manufactured by
ASTELCO Systems\footnote{\url{http://www.astelco.com}}.  With an
alt-azimuthal mount, the image of the sky rotates in the focal plane
during the time of data acquisition then a derotator is needed to
capture long-exposure images.
The optical scheme comprises a primary concave parabolic, $0.8\meter$
mirror $M1$ made of Schott Zerodur $85\milli\meter$ height, and coated
with Al+MgF$_2$ with reflectivity greater than 95\%, which reflects
light towards a secondary convex hyperbolic mirror $M2$, also used for
focusing; a comparatively small tertiary flat mirror $M3$ reflects the
light to one of the two Nasmyth foci of the telescope at $f/8$, with a
focal length of $6.4\meter$.

In fact, a peculiarity of \oarpaf consists of the tertiary flat
mirror that can be manually rotated using a handle to switch
the Nasmyth focus between:
\begin{itemize}
\item the ``observing flange'', for outreach usage, provided with a
  manual focuser and a set of oculars;
\item the ``scientific flange'', where a field derotator is placed.
  The scientific flange includes a field flattener, in order to
  flatten the focal plane so that not only the central point but the
  whole image is fully focused.
\end{itemize}
Up to now, we employed the telescope for commissioning and observation
using the \stl (gray box in Fig.~\ref{fig:tel}) on the scientific
flange, and we measure a plate scale of $0.29\arcsecond / \pixel$.

%%%%%%%%%%%%%%%%%%%%%%%%%%%%%%%%%%%%%%%%%%%%%%%%%%%%%%%%%%%%%%%%%%%%
\subsection{Pointing model}
\label{sec:point-model-telesc}

The telescope can move at a velocity of $20 \degree / \second$, and an
acceleration of $20 \degree / \second^2$ in order to point a field.
Accurate pointing or positioning of a telescope is of paramount
importance for any telescopic system in order to be productive. The
telescope has various static and dynamic pointing errors: these must
be compensated with the help of position measurements of reference
stars.  The \textsc{AsTelOS} proprietary software released by
Astelco Systems allows users to make a new pointing model or improve
an existing one by adding more measurements.  The software algorithm
comprises 25 coefficients, computed by measuring the offset
between the instrumental position of an object and its calculated
theoretical position.
The pointing model was performed with about 50 stars; in the past
years, since only few observations were performed due to the
impossibility to operate remotely, the mechanics of the telescope were
little solicited and the model proved to be stable on a range of
several months. However, when the full remote control will be completed,
hopefully by 2021, the duty cycle will also largely increase and we
plan to redo the pointing model once per week (by remote, at that
point).
After refining our pointing model, we reached a pointing accuracy
$<10\arcsecond$ root mean square. Night tests showed that during a 30
minutes observation on the same target, the tracking precision is
$<1\arcsecond$.
%

%%%%%%%%%%%%%%%%%%%%%%%%%%%%%%%%%%%%%%%%%%%%%%%%%%%%%%%%%%%%%%%%%%%%
\section{Site characterization}
\label{sec:site}

To allow for scientific measurements and to verify the feasibility of
new ideas, it is important to characterize the site. The main
measurements and calibrations have been performed using the \stl and
include:
\begin{inparaenum}
\item Determination of the average sky background;
\item Determination of the typical seeing of the site;
\item Determination of the extinction coefficients and zero points.
\end{inparaenum}

%%%%%%%%%%%%%%%%%%%%%%%%%%%%%%%%%%%%%%%%%%%%%%%%%%%%%%%%%%%%%%%%%%%%
\subsection{Average sky background}
\label{sec:aver-sky-backgr}

The largest contribution to the sky emission comes from the
moonlight\cite{Cramer2013, 2014Msngr.156...31J}, which peaks
  at around $550\nano\meter$ and decreases towards larger wavelengths
becoming negligible in the near IR. Of course, the scattered moonlight
 strongly depends on the moon phase and changes the sky brightness by
huge factors, up to 30 in the visible band.
We used the \stl at \oarpaf to measure a typical sky brightness during
a new moon phase. We find a value of of $22.40\, m_{AB}/ \arcsecond^2$
in the $B$ filter, down to $21.14\, m_{AB}/ \arcsecond^2$ in the $I$
filter.

%%%%%%%%%%%%%%%%%%%%%%%%%%%%%%%%%%%%%%%%%%%%%%%%%%%%%%%%%%%%%%%%%%%%
\subsection{Seeing}
\label{sec:seeing}

The distortion induced by the atmosphere is expressed by the seeing
parameter.  An estimate of the typical seeing at the \oarpaf site was
measured by determining the point spread functions of several stars,
assumed to be point sources, and fitting with 2-dimensional Gaussian
functions. The seeing is the average Full Width at Half Maximum
(FWHM).
During a typical summer night, when the seeing is expected to be the
worst because of the warm conditions, we find at \oarpaf
using the \stl a seeing of $2.5\arcsecond$ and a general
range of variability of between $1.5\arcsecond$ and $3.0\arcsecond$
during the year.

%%%%%%%%%%%%%%%%%%%%%%%%%%%%%%%%%%%%%%%%%%%%%%%%%%%%%%%%%%%%%%%%%%%%
\subsection{Extinction and zero point}
\label{sec:phot-calibr}

% ----------------------------------------------------------------
\begin{figure}[t]
  \centering
  \includegraphics[width=\columnwidth]{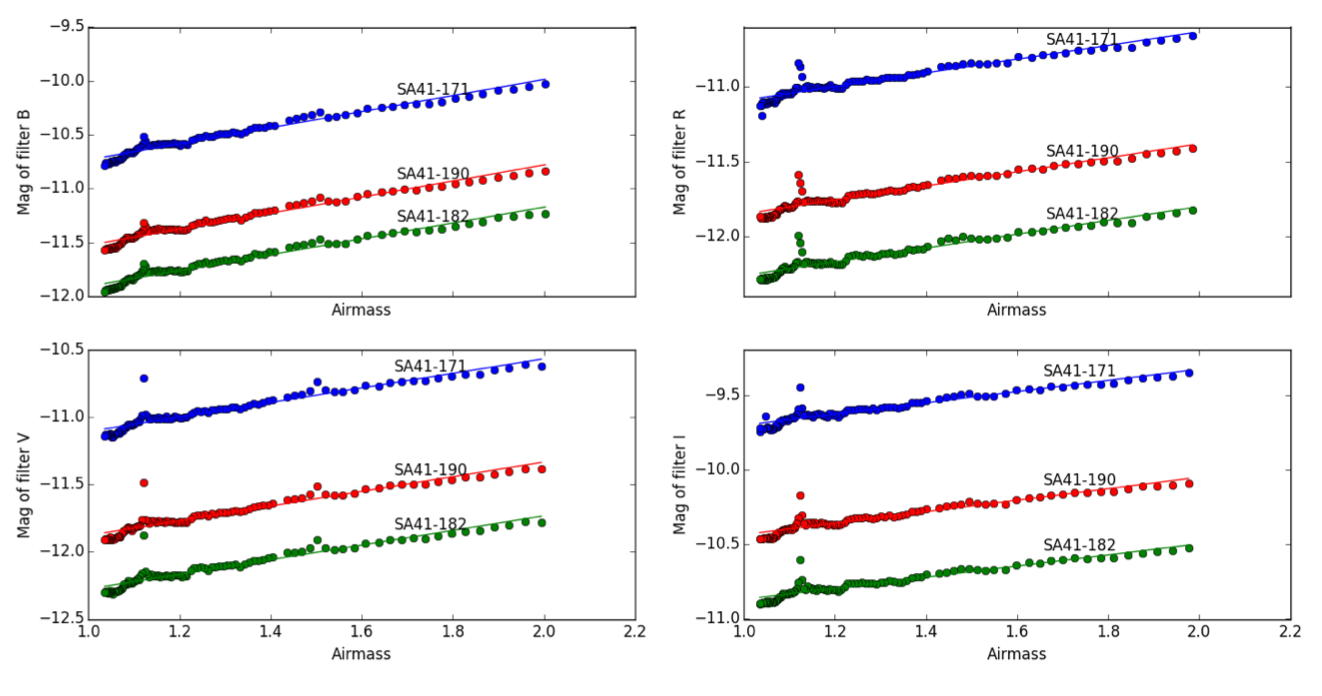}
  \caption{Magnitude as a function of the air mass obtained from three
    reference stars whose light was captured by the \stl camera using
    several filters. Data derived from sequences of $BVRI$
      images. We attribute features in the curves, such as the ones at
      an airmass of 1.1, to the change in the sky condition during one
      of the sequences.  }
  \label{fig:filt}
\end{figure}
% ----------------------------------------------------------------

When an image is acquired, the conversion factor from counts to
scientific units is not known a priori and a photometric calibration
of all instruments is mandatory~\cite{Stetson1989,
  2000hccd.book.....H}. In this respect, a series of standard stars,
whose light output in various passbands of photometric systems has
been carefully measured and selected by the TOPCAT program
\cite{2005ASPC..347...29T}, are used. The stars are observed at their
maximum elevation in the sky, typically at an air mass value ($AM$)
smaller than 1.15.  The instrumental zero points are determined as a
function of the color index in different passband filters. The
calibrated magnitude $M_\lambda{_1}$ at a defined passband centered
around the wavelength $\lambda_1$ is the sum of more components:
$M_\lambda{_1} = z_\lambda{_1} + m_\lambda{_1} - k_\lambda{_1} AM +
c_\lambda{_1} (m_\lambda{_1} - m_\lambda{_2})$, where
\begin{inparaitem}[]
\item $z_\lambda{_1}$ is the zero point of the photometric
  system at a defined passband,
\item $m_\lambda{_1}$ is the instrumental magnitude,
\item $k_\lambda{_1}$ is the atmospheric extinction coefficient,
\item $AM$ is the air mass at the observation time,
\item $c_\lambda{_1}$ is a color term and the difference
\item $m_\lambda{_1} - m_\lambda{_2}$ is the instrumental color index
  from two different filters.
\end{inparaitem}
In fact, the atmospheric extinction is a complex phenomenon to model
because many effects are involved; it is more prominent in the $U$,
$B$ and $V$ filters, whereas it is much smaller in the $R$ and $I$
filters.
As a first approximation, the extinction has a first order term
proportional to the air mass at the time of observation, which takes
into account the attenuation due to the mass of air traversed by
photons, and a second order term, which takes into account its
influence on the color variation.
At the effective wavelength $\lambda$, the instrumental magnitude $m$
is related to the extra-atmosphere instrumental magnitude $m_0$ by
Bouguer's law $m = m_0 + k_\lambda AM$, where $k_\lambda$ is the
extinction coefficient (measured in mag/$AM$).  A fit of observed
magnitudes of standard stars at different air masses allowed
determining the extinction coefficients for every filter.

Results of extinction coefficient at \oarpaf, calculated using the
\stl and the related filters, are shown in Table~\ref{tab:filt} and
Fig.~\ref{fig:filt}. We will repeat these operations
  with the \stx.
Once these have been determined, the zero points and the color terms
can be in turn measured. We report the results in Table~\ref{tab:zero}.
We verified the quality of the determination of the zero points and
color terms by comparing the theoretical expected magnitudes with the
$M_\lambda$ calculated magnitudes for various known sources.

% ----------------------------------------------------------------
\begin{table}[b]
  \centering
  \subfloat[ Zero points and color term.]
    {
  \begin{tabular}{ccr}
    \hline
    $\lambda_1$ ($\lambda_2$) &  Zero point     & Color term      \\
    \hline
    \hline
    B (V)   & $23.03\pm 0.06$ & $-0.26\pm 0.12$ \\
    V (B)   & $22.72\pm 0.03$ &  $0.04\pm 0.06$ \\
    V (R)   & $22.73\pm 0.02$ &  $0.14\pm 0.13$ \\
    R (V)   & $22.40\pm 0.04$ &  $0.60\pm 0.40$ \\
    R (I)   & $22.63\pm 0.25$ &  $0.15\pm 0.19$ \\
    I (R)   & $21.93\pm 0.40$ &  $0.90\pm 0.30$ \\
    \hline
  \end{tabular}
  \label{tab:filt}}
  \qquad
  \subfloat[Measured extinction coefficients for each passband filter.]
    {
  \begin{tabular}{cc}
    \hline
    Passband filter & Extinction coefficient \\
    \hline
    \hline
       B            & $0.742\pm 0.005$ \\
       V            & $0.543\pm 0.002$ \\
       R            & $0.463\pm 0.005$ \\
       I            & $0.376\pm 0.005$ \\
    \hline
  \end{tabular}
  \label{tab:zero}}
\caption{ Site extinction coefficients, color terms and
    instrumental zero points measured with the \stl.  Color terms are
  determined for wavelength $\lambda_1$ at the nominal central value
  of each filter, with respect to a $\lambda_2$ at the value of a
  nearby filter.}
\end{table}
% ----------------------------------------------------------------

%%%%%%%%%%%%%%%%%%%%%%%%%%%%%%%%%%%%%%%%%%%%%%%%%%%%%%%%%%%%%%%%%%%%
\section{Instruments commissioning}
\label{sec:commissioning}

During the first phase of the observatory life, we commissioned and
employed only the \stl camera. In particular, main measurements and
calibrations that were performed include:
\begin{inparaenum}
\item Measurement of the dark current of the CCD;
\item Determination of the quantum efficiency of the CCD;
\item Determination of the efficiency of the photometric filters.
\end{inparaenum}
Using the aforementioned information, an Exposure Time Calculator
(ETC) was written. The ETC allows us to determine the needed
exposure time for observing a given target, based on the desired
signal-to-noise ratio, considering the photon flux of the source, all
attenuation and distortion factors, and all background sources.

%%%%%%%%%%%%%%%%%%%%%%%%%%%%%%%%%%%%%%%%%%%%%%%%%%%%%%%%%%%%%%%%%%%%
\subsection{Detectors and filters}
\label{sec:ccds-phot-filt}

% % ----------------------------------------------------------------
% \begin{figure}[t]
%   \centering
%   \includegraphics[width=0.5\columnwidth]{fig1}
%   \caption{Determination of the electron signal from dark noise as a
%     function of the exposure time at -6 \degreeCelsius. Results of a
%     linear fit to the data are superimposed. }
%   \label{fig:stl}
% \end{figure}
% % ----------------------------------------------------------------

% ----------------------------------------------------------------
\begin{figure}[t]
  \centering
  \includegraphics[width=\columnwidth]{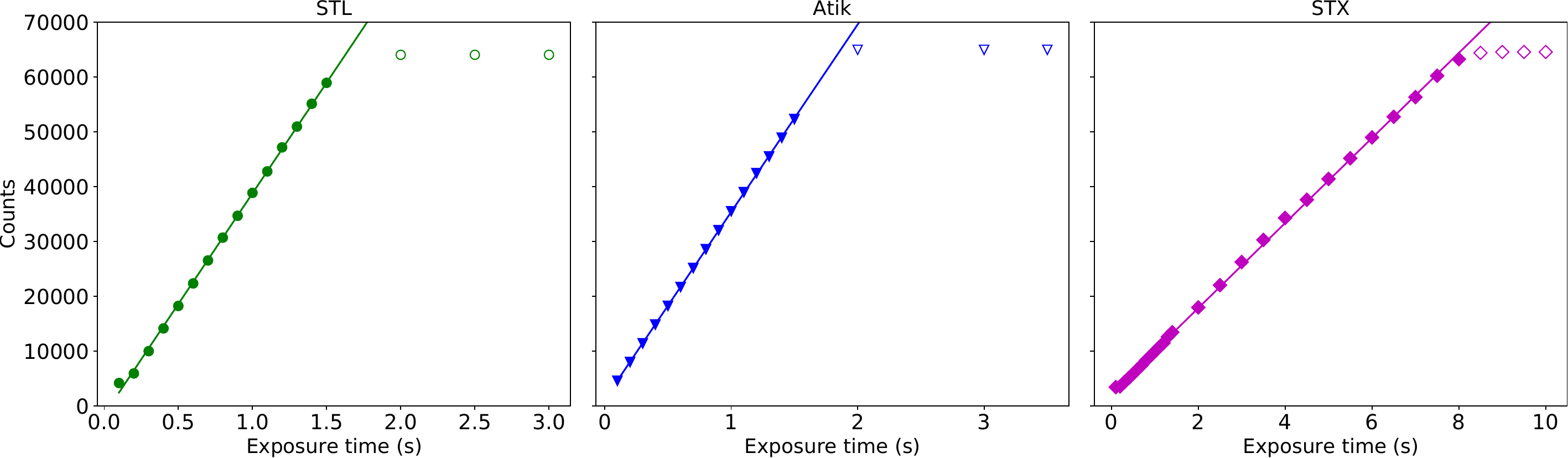}
  \caption{Linearity test of the \stl, the \atik, and the \stx at
    -10\degreeCelsius, +1\degreeCelsius, and -20\degreeCelsius,
    respectively. Results of a linear fit are superimposed to the data
    points. Empty dots have been sigma-clipped. All cameras are linear
    from approximately 4\,000 counts to saturation level.}
  \label{fig:stx}
\end{figure}
% ----------------------------------------------------------------

\oarpaf originally had two CCDs: a \stl and a \atik.  Both use a Kodak
\textsc{kai-11000} sensor with 11 million
$9\,\micro\meter \times 9\,\micro\meter$ pixels, covering
$36 \times 24.7 \,\milli\meter$. In 2019, a new \stx with a Class-1
CCD was purchased. This camera adopts a Kodak \textsc{kaf-16801} sensor
with 16 million $9\micro\meter \times 9\micro\meter$ pixels, covering
$36.8 \times 36.8 \,\milli\meter$.  All three CCDs are equipped with
Peltier coolers and can reach at least $20$--$30\degreeCelsius$ below
the ambient temperature.

We measured the plate scale of the \stl on the observing
  flange of the telescope to be $0.29\,\arcsecond / \pixel$, for a
  Field of View of around $20\,\arcmin$ on the long side
  ($\approx 4000$px). We foresee the same results to hold for the
  other two cameras due to the fact that the three cameras have the
  same pixel size.  ADCs use 16 bits for the readout of the detectors,
  the \underline{readout time} is $20$--$30\,\second$ for the \stl and the
  \atik, down to $12\,\second$ for the new \stx, and stored in FITS
  format by means of the software released by the manufacturer.

The quantum efficiency of the CCDs, defined as the ratio
between incoming photons to converted electrons, depends on the
wavelength, with a maximum of around 50\% at $500\,\nano\meter$.

The photometric filters are a set of standard $UBVRI$ Johnson-Cousins
filters~\cite{1990PASP..102.1181B}, centered around the wavelengths
360 ($U$), 420 ($B$), 550 ($V$), 640 ($R$) and 790 ($I$) \nano\meter,
respectively, with a full width at half maximum (FWHM) bandwidth of
60, 90, 85, 140 and 150 \nano\meter, respectively.

The readout noise was measured exploiting the dedicated software
  MaxIm\footnote{\url{https://diffractionlimited.com/product/maxim-dl/}}
  and it was found to be consistent with the declared noise at
  construction, of around $11 e^-$.

%%% Linearity, Gain, Dark current
The existence of a linear relation between the charge collected within
each pixel and the digital number stored in the output image was
verified with exposures of various duration on a uniform and constant
light source: a linear relation holds up to very high counts, where a
deviation from linearity is seen. Fig.~\ref{fig:stx} shows the counts
in ADU as a function of the exposure time for the three CCDs.  We
found a good linearity range for all detectors (4\,000-60\,000 counts)
with a gain $G$ of
\begin{itemize}
\item $1.040 e^-/\rm ADU$ for the \stl,
\item $0.996 e^-/\rm ADU$ for the \atik, and
\item $0.998 e^-/\rm ADU$ for the \stx.
\end{itemize}
The dark current $D$ was determined by means of sets of $60\minute$
exposure dark frames at the lowest temperature that each
  camera could reach. We obtain:
\begin{itemize}
\item $0.70 \rm ADU/\pixel/\second$ for the \stl at $-10\degreeCelsius$,
\item $0.94 \rm ADU/\pixel/\second$ for the \atik, at $-3\degreeCelsius$, and
\item $0.28 \rm ADU/\pixel/\second$ for the \stx, at $-20\degreeCelsius$.
\end{itemize}
In particular, \underline{we also obtain the value of
  $0.70 \rm ADU/\pixel/\second$ for the \stx at $-10\degreeCelsius$,}
while the temperature of the \atik suggests a potential issue with the
Peltier cells, which will be the object of further investigations.

% \stl
% T = 0°C -> 1.09
% T = -5°C -> 0.87
% T = -10°C -> 0.70

% \atik
% T = 1°C -> 0.094
% T = 0°C -> 0.097
% T = -3°C -> 0.094

% \stx
% T = -10°C -> 0.29
% T = -15°C -> 0.28
% T = -20°C -> 0.28

Having now three CCDs, we plan to use them as follows:

\begin{itemize}
\item The \stl will be switched to the new \lhires spectrograph;
\item The \atik is used in conjunction with the \flechas spectrograph.
\item The new \stx will be dedicated to high-performance photometry,
  replacing the \stl which has been used until now, due to the
    fact that we want to employ a Class A, full-frame CCD for the main
    use of the telescope, i.e. photometric observations. Furthermore,
    we could pair the new camera with a on-axis STX-Guider and a
    tip-tilt corrector.
\end{itemize}

A preliminary study aiming at pairing these three components
  in a three-headed instrument\cite{2020SPIE11447E..5JC} has recently
  been presented at an international conference: flat mirrors on a linear
  stage will allow us to select among the photometric head, the long
  slit spectroscopy head, and the échelle spectroscopy head.

%%%%%%%%%%%%%%%%%%%%%%%%%%%%%%%%%%%%%%%%%%%%%%%%%%%%%%%%%%%%%%%%%%%%
\subsection{The \flechas spectrograph}
\label{sec:flechas}

% ----------------------------------------------------------------
\begin{figure}[t]
  \centering
  \includegraphics[width=0.85\columnwidth]{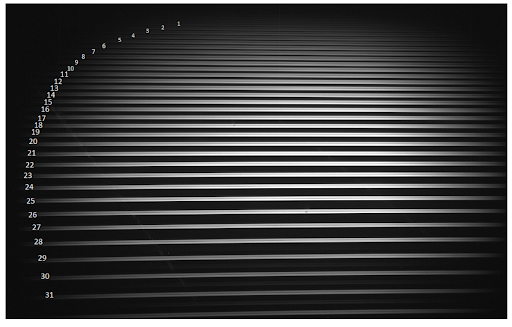}
  \caption{\flechas spectrum taken with the \atik. }
  \label{fig:flechas}
\end{figure}
% ----------------------------------------------------------------

% ----------------------------------------------------------------
\begin{figure}[t]
  \centering
  \includegraphics[width=0.99\columnwidth]{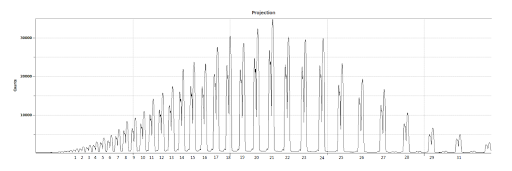}
  \caption{Trend of the counts in the cross direction to the
dispersion for the 31 orders of the \flechas. }
  \label{fig:peak}
\end{figure}
% ----------------------------------------------------------------

\oarpaf is equipped with \flechas, a Fiber-Linked ECHelle Astronomical
Spectrograph~\cite{2014AN....335..417M}: it is an échelle spectrograph
specifically designed for class $1\meter$ telescopes with focal ratios
of $f/8$ to $f/12$ of the Astelco Systems company.  The optical design
is optimized for seeing conditions of around $1.5\arcsecond$ and for
typical pixel sizes of most common CCDs. The expected seeing spot size
determines a pin hole size of around $150 \micro\meter$ and, in turn,
yields an effective resolving power $R\sim 9300$, increased to
  $R\sim 15000$ with an image slicer to suppress scintillation
  effects; here $R$ is defined as the ratio $\lambda / \Delta\lambda$
between a given wavelength $\lambda$ and the minimum resolvable
wavelength difference $\Delta\lambda$.

\begin{figure}[t]
  \centering
  \includegraphics[width=0.49\columnwidth]{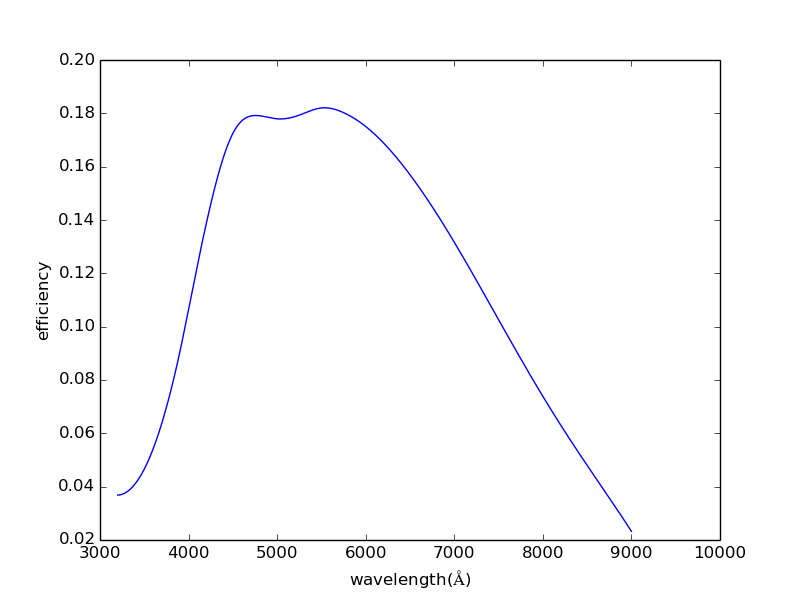}
  \caption{ Overall efficiency of the \flechas. }
  \label{fig:efficiency}
\end{figure}

The wavelength range of the spectrograph optics is
$350$--$850\,\nano\meter$, We paired the \flechas with the
  \atik. The overall efficiency of the spectrograph and the camera is
  shown in Fig.~\ref{fig:efficiency}. The maximum of efficiency is
  $\sim 18.3\%$ at 555 \nano\meter .

% ----------------------------------------------------------------
% ----------------------------------------------------------------

We measured the stability of \flechas using an embedded Thorium-Argon
(ThAr) calibration source and we find to be $0.46\pixel$, at 95\%
confidence level over one hour.

The \flechas equipped with the \atik covers 31 échelle orders, as seen
in Fig.~\ref{fig:flechas}. We measured a variable order separation
from $4\pixel$ to $105\pixel$. All orders measure about $30\pixel$ in
height, exhibiting a double peak profile because of the image slicer
(see~Fig.~\ref{fig:peak}).

The calibration in wavelength and the dispersion were obtained again
by means of the ThAr lamp. The dispersion $n$, expressed in
$\nano\meter/\pixel$, increases with wavelength with a linear relation
of the form: $n = a\lambda + b$, with $\lambda$ expressed in
$\nano\meter$.
We found $a$ and $b$ to be
$1.39\times 10^{-6}\, / \pixel$ and
$1.45\times 10^{-4}\, \nano\meter / \pixel$, respectively.
%(see~Fig.~\ref{fig:orders}).

\begin{figure}[t]
  \centering
  \includegraphics[width=0.49\columnwidth]{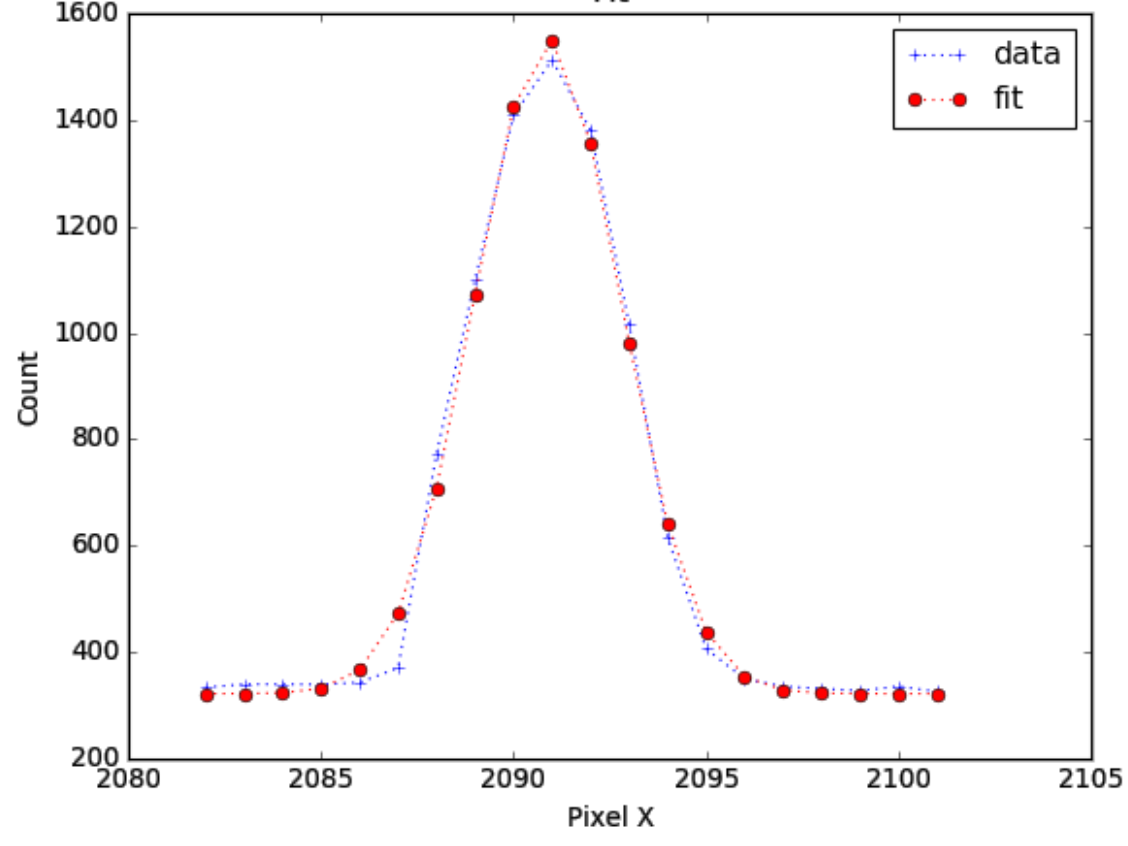}
  \caption{An example of Gaussian fit (red dots) to a
     ThAr peak data (blue crosses): the FWHM of this peak
     is found to be $4.46\pixel$.  }
  \label{fig:thar}
\end{figure}

\begin{figure}[t]
  \centering
  \includegraphics[width=0.49\columnwidth]{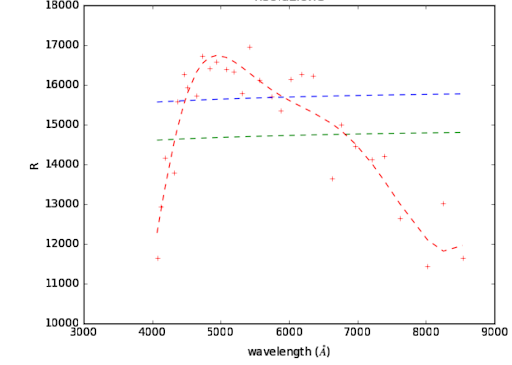}
  \caption{Resolving power $R$ as a function of the
     wavelength (red). The blue (green) lines use the
     median (mean) FWHM.  }
  \label{fig:reso}
\end{figure}

The spectral resolving power of \flechas as a function of $\lambda$
was determined by a Gaussian fit to the peak along the direction of
dispersion (see~Fig.~\ref{fig:thar}), in order to measure the FWHM of
the emission lines detected in the ThAr calibration spectra and
averaging the FWHM over the various orders (Figs.~\ref{fig:thar}
and~\ref{fig:reso}).

% Such measured resolving power allows us to
% establish\cite{2001A&A...374..733B} that the minimum appreciable
% radial velocity of a star is around $15.8 \kilo\meter / \second$.

% ----------------------------------------------------------------
% ----------------------------------------------------------------

% ----------------------------------------------------------------
% ----------------------------------------------------------------
% % ----------------------------------------------------------------
% \begin{figure}[t]
%   \centering
%   \subfloat[An example of Gaussian fit (red dots) to a
%      ThAr peak data (blue crosses): the FWHM of this peak
%      is found to be $4.46\pixel$.]
%      {\includegraphics[width=0.47\columnwidth]{gausfit}
%        \label{fig:thar}
%      } \quad
%   \subfloat[Resolving power $R$ as a function of the
%      wavelength (red). The blue (green) lines use the
%      median (mean) FWHM.]
%      {\includegraphics[width=0.49\columnwidth]{Rpower}
%      \label{fig:reso}}
%    \caption{Example of ThAr peak data and resolving power of the
%      \flechas.}
% \end{figure}
% % ----------------------------------------------------------------

%%%%%%%%%%%%%%%%%%%%%%%%%%%%%%%%%%%%%%%%%%%%%%%%%%%%%%%%%%%%%%%%%%%%

\subsection{The \lhires spectrograph}
\label{sec:lhires}

In 2019, a Shelyak \lhires medium-resolution was purchased. This is a
long-slit spectrograph with slit width of $25 \micro\meter$, 1200
grooves per \milli\meter\ grism and a resolving power of
$R\sim 5800$ according to the vendor, which will be verified in
  further tests, as well as the stated total efficiency of $\sim 7.3\%$.

We decided to pair the \lhires to the \stl.
In this configuration, and based on the producer's documentation, we
estimate at \oarpaf a maximum observable magnitude of $m_V\approx 10$,
for a signal-to-noise ratio of $100$ and 1 hour exposure. We also
calculate a mean sampling of about $0.034 \,\nano\meter / \pixel$,
which in turn yields a mean field width of one
acquisition\cite{Ricci2020} of $\approx 138 \,\nano\meter$.
A micro-metric screw allows the shift of the central wavelength, so
that most of the $\approx 300\nano\meter$ wavelength range of the \stl
can be in principle covered with two exposures. The Shelyak
spectrograph has not been fully commissioned yet: its calibration and
first usage will be presented in a different publication.
We foresee the slit length on the sky to have the same size of
  the photometric field, so $\approx 20\arcmin$.

%%%%%%%%%%%%%%%%%%%%%%%%%%%%%%%%%%%%%%%%%%%%%%%%%%%%%%%%%%%%%%%%%%%%
\section{\underline{Calibrations}}
\label{sec:calibration}

% ----------------------------------------------------------------
\begin{figure}[t]
  \centering
  \includegraphics[width=\columnwidth]{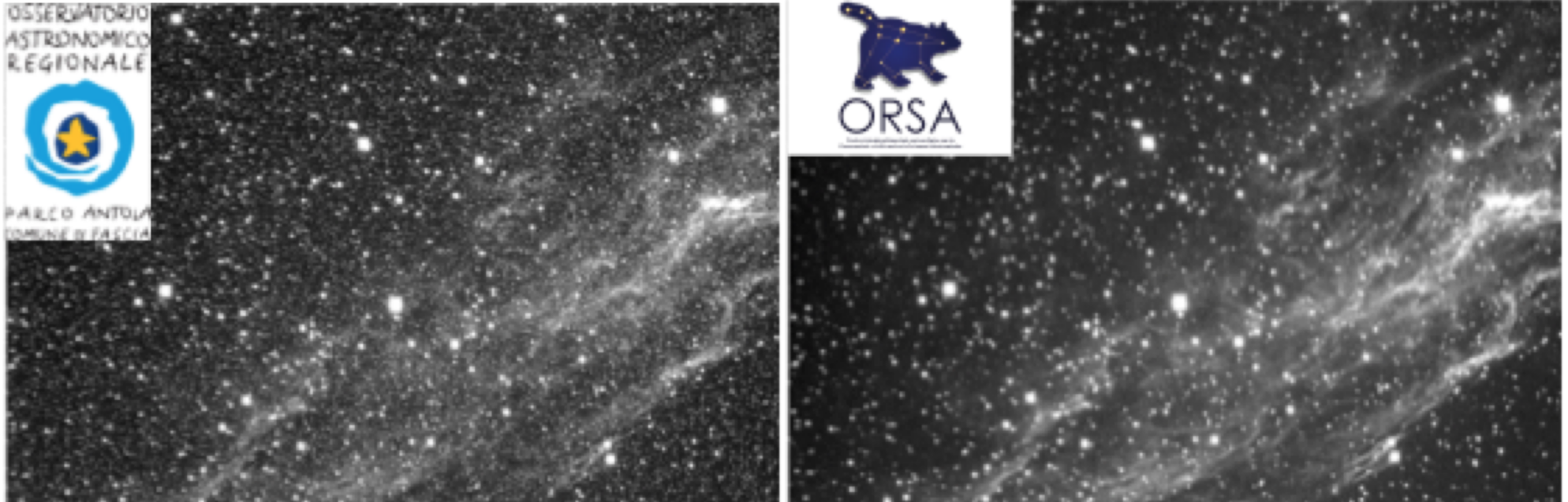}
  \caption{The telescope and a sample image of the Velo Nebula (before
    and after debiasing and flat-fielding used for outreach purposes.}
  \label{fig:vela}
\end{figure}
% ----------------------------------------------------------------

The calibration procedure is mainly intended to remove additive
contributions to the background, such as the electronic pedestal
level, the dark current, the multiplicative gain, and illumination
variations across the chip. The goal of the data reduction pipeline
is ideally to remove signatures of experimental distortions
  from the data, thus allowing us to achieve the most accurate values for
  the observable and to  minimize the contribution of deterministic
factors in the uncertainties and, at the same time, by preserving
information about noise sources, so that users can evaluate the
random uncertainties of the reduced data. The raw counts of a pixel in
position $x,y$ in a CCD frame can be computed as
$s(x,y)=B(x,y)+tD(x,y)+tG(x,y)I(x,y)+N$, where
$B(x,y)$ is the bias value of each pixel,
$t$ is the integration time,
$D(x,y)$ is the dark current, here expressed in ADU/pixel/s,
$G(x,y)$ is the sensitivity gain and
$I(x,y)$ is the light flux reaching the pixel including the signal from the source as well as the sky background, while
\underline{$N(x,y)$ is the readout noise and any other irreducible noise source of the pixels, to include information on dead or hot pixels.}
$B$, $D$ and $G$ are measured from bias, dark and flat frames,
respectively.

Bias frames measure the readout noise and correspond to observations
without exposure to light (shutter closed) for a total integration
time of 0 seconds: several frames are acquired so that a median frame can be
determined with reduced statistical uncertainty. This master
  frame is subtracted from data during reduction. If we allow the CCD
to integrate for some amount of time, without any light falling on it,
there will be a signal caused by thermal excitation of electrons in
the CCD: this is called dark signal and it is very sensitive to
temperature.

All CCDs have non-uniformities, that is, a uniform illumination of the
CCD does not yield an equal signal in each pixel (even ignoring
noise). Small scale (pixel to pixel) non-uniformities (typically a few
percent from one pixel to the next) are caused by slight differences
in pixel sizes.  Larger scale (over large fractions of the chip) non
uniformities are caused by various effects, such as small variations
in the silicon thickness across the chip, non-uniform illumination
caused by telescope optics (vignetting). These can sum up to
variations of around 10\% over the chip. To measure and correct for
these non-uniformities, the entire CCD is illuminated by a uniform
source of light and flat field data are taken.
\underline{At \oarpaf, the twilight method~\cite{1993AJ....105.1206T}
  has been applied in order to acquire sky flats for photometry.  The
  new dome also provides a setup for dome flat fields, which will be
  used as a backup solution.}

%%%%%%%%%%%%%%%%%%%%%%%%%%%%%%%%%%%%%%%%%%%%%%%%%%%%%%%%%%%%%%%%%%%%
\section{Data reduction pipeline}
\label{sec:pipeline}

A Python pipeline has been developed to remove the various background
sources and derive a pre-treated image, as shown in
Fig.~\ref{fig:vela}.  The development of the pipeline also foresees
field solving by querying \url{astrometry.net}, and matching sources
from the \texttt{GAIA DR2} catalog~\cite{2016A&A...595A...1G,
  2018A&A...616A...1G} in order to perform automatic aperture
photometry.  Field solving also overcomes errors due to the
possibility of low tracking accuracy, especially on derotation. If
this happens, light from a given object will not populate the same
pixel for each frame. Further implementation foresees the use of a
separate, on-axis guiding camera and the tip-tilt lens corrector in
order to correct the pointing during frame acquisition.

Currently, data reduction pipeline is still under
  development. It produces light curves of stellar fields, in the form
  of ascii output, (tested on several exoplanetary transit targets) in
  fully-automatic mode, as well as complete pre-treated,
  astrometric-resolved fields in FITS format, containing a large
  number of derived parameters in the FITS Header, following a
  ``ESO-like'' standard. Long slit and échelle spectroscopy will be
  the next goals.

%%%%%%%%%%%%%%%%%%%%%%%%%%%%%%%%%%%%%%%%%%%%%%%%%%%%%%%%%%%%%%%%%%%%
\section{Remote control}
\label{sec:remote-control}

Initially, the facility and the dome of the telescope were conceived
to be operated only locally and manually.  Consequently, the
scientific usage of the telescope was limited.  Of course, being able
to remotely control the dome and the other instruments represents a
huge advantage for the scientific exploitation of the
telescope. Therefore, with the goal of remotely controlling it, we
implemented a first modification towards an automated use of the dome
by developing a Python script running on a Raspberry-PI to query the
control center for position, and interacting with the dome by an
Arduino platform.  The necessary works to remotely control the dome
started immediately, and the needed set of instruments was bought:
these include a new weather station with rain-gauge, anemometer,
hygrometer and thermometer, an all-sky camera, two IP webcams, and a
new electronic system with an encoder directly interfaced to the
interlock system.

This way, remote operators can monitor the weather and sky
conditions at \oarpaf and the possible presence of visitors; for the
safety of the telescope, the dome will automatically close in case of
bad weather and in case of prolonged absence of the Internet
connection.  A custom software framework under the Linux operating
system was designed to manage all the parts.

Since remote operations must be fully reliable, a stable access to the
Internet is mandatory.  Unfortunately, the access link currently in use
at \oarpaf does not yet satisfy the necessary requirements of
stability. Several actions for improving the overall
quality-of-service and the service level agreements of the Internet
connection are planned during 2021 by the Liguria Digitale
company\footnote{\url{https://www.liguriadigitale.it/}}.  In the near
future, the remote control of the facility will be obtained in two
ways, described hereafter.

%%%%%%%%%%%%%%%%%%%%%%%%%%%%%%%%%%%%%%%%%%%%%%%%%%%%%%%%%%%%%%%%%%%%
\subsection{Ricerca}
\label{sec:ricerca}

To date, a full remote control is not yet available. The newly
installed dome was provided with a commercial control software, named
\textit{Ricerca}, developed by
OmegaLab\footnote{\url{http://atcr.altervista.org/ita/index.html}},
operating on Windows. \textit{Ricerca} makes use of ASCOM drivers and
can be used together with the telescope software to remotely manage
the observatory and to monitor the system, the dome, and the
detectors. The software interfaces with a hardware module called
\texttt{OCSIII}, provided with relay control switches for the several
mentioned subsystems, also including a screen and dome light
  for flat fields, which may be used for calibrations along with sky
  flat fields. This first remotization step will be completed before
  summer 2021, after the purchase of the required software licenses.

%%%%%%%%%%%%%%%%%%%%%%%%%%%%%%%%%%%%%%%%%%%%%%%%%%%%%%%%%%%%%%%%%%%%
\subsection{Web interface}
\label{sec:web-interface}

To allow more flexibility in the framework for future development, a
completely Linux-based custom control software is being set up.  The
latter uses top modern internet technologies to operate through
commands given via a web browser, or script, through a web Application
Programming Interfaces (API).

The development ``from scratch'' of the remotization framework for a
recent observatory can involve modern solutions that \underline{do not
  need to rely on decades of hardware/software substrates}. For this
reason, we are considering to implement the front end interface using
\texttt{html} and \texttt{javascript} languages via the popular
\texttt{Bootstrap 4} framework, whereas the V8-based technology
\texttt{node.js} and \texttt{mongodb} are used for the server side
development and storing purposes~\cite{Ricci2020}.  It already allows
the control of the \textsc{sbig} cameras and the all-sky camera as a
test benchmark and it will be implemented in order to monitor and
control the dome and the telescope, as well as the weather station and
the webcams, producing real-time images.

Moreover, it will allow controlling and scheduling public visits and
scientific operations. In addition, it will be easily usable by
schools for public events and outreach activities.

%%%%%%%%%%%%%%%%%%%%%%%%%%%%%%%%%%%%%%%%%%%%%%%%%%%%%%%%%%%%%%%%%%%%
\section{Science cases}
\label{sec:science}

The scientific reach of \oarpaf is wide. Observations and preliminary
results achieved during the past years demonstrate the large potential
of the facility: these include observations for which small,
$1\,\meter$-class telescopes, such as \oarpaf, have enough sensitivity
to be competitive for measurements that require long campaigns,
typically not possible with big telescopes, whose observation time has
to be shared among a number of projects.

The scientific potential of the telescope has been presented in
several
national\footnote{\url{http://events.iasfbo.inaf.it/gloria/od_programme.php}}
and international conferences~\cite{2016NCimC..39..284R, Cabona2016a,
  Ricci2016, Ricci2020} and in master-degree theses of students of the
University of Genova~\cite{Righi2015, Cabona2016, Nicolosi2019}.

In the following subsections we present such results and future
prospects.

%%%%%%%%%%%%%%%%%%%%%%%%%%%%%%%%%%%%%%%%%%%%%%%%%%%%%%%%%%%%%%%%%%%%
\subsection{Exoplanetary transits}
\label{sec:exo-planet-transits}

% ----------------------------------------------------------------
\begin{figure}[t] \centering
  \includegraphics[width=\columnwidth]{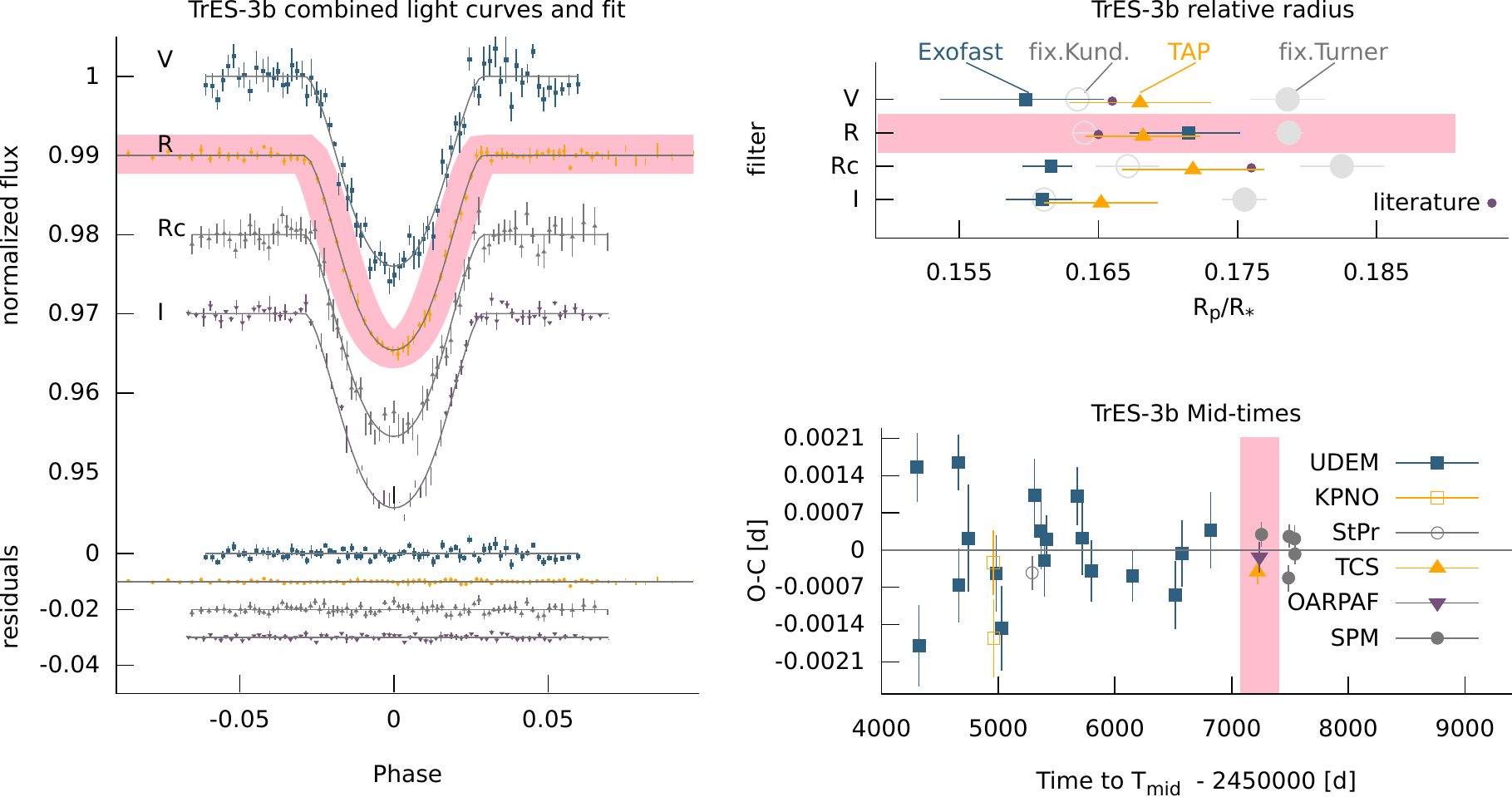}
  \caption{\texttt{TrES-3b} recent results\cite{2017PASP..129f4401R}
    obtained with \oarpaf contribution (red marks).  Left: Light
    curves fitting and residuals. Error bars are around 2--5mmag
    peak-to-valley in all filters.  Top: Ratio between \texttt{TrES-3b}
    and host star radii obtained with different procedures (EXOFAST
    and TAP modelling), compared with values found in the literature
    (grey). Bottom: Observed-Calculated Mid-Time transits vs Mid-times.
  }
  \label{fig:tres}
\end{figure}
% ----------------------------------------------------------------

Planets orbiting stars other than the Sun have been first discovered
in the 90s and along the years their number increased and passed
4000\footnote{\url{http://exoplanet.eu}}. They can be detected with
different experimental methods. For \oarpaf, the photometric transit
\cite{2000ApJ...529L..45C} is the most suitable method. To measure the
needed light curves, we adopted the defocused photometry
\cite{2013MNRAS.434.1300S} allowing the star hosting the target
exoplanet to cover many pixels, and to obtain magnitude dispersion
levels that have been proved to be  comparable to those of
space telescopes in observatories of the same class of
  \oarpaf, such as GROND or the Danish $1.54\meter$. The measurements
thus obtained are practically unaffected by any instrumental defects,
by calibration biases, and by variations of the quality of the
observing night, an important aspect for measurements that can
possibly last several hours.  These results are indeed within
  the range of \oarpaf, but further instrument setup issues, such as
  the star chasing or a tip tilt correction, have to be implemented.

At \oarpaf, we observed, with the \stl, the exoplanet \texttt{TrES-3b}
using the defocused photometry technique. The observation contributed
to a peer reviewed publication~\cite{2017PASP..129f4401R}, where data
also gathered at the Observatorio Astronomico Nacional de San Pedro
Martir (Mexico), Observatorio de la Universidad de Monterrey (Mexico)
and Telescopio Carlos Sanchez at the Observatorio del Teide (Spain)
were combined (Fig.~\ref{fig:tres}), thus allowing us to derive
physical and orbital parameters of the planet. Other exoplanets that
we observed include \texttt{WASP-58b}, \texttt{HAT-P-3b} and
\texttt{HAT-P-12b} \cite{Cabona2016, Cabona2016a}: the related data
reduction is ongoing using the pipeline described in
Sect.~\ref{sec:pipeline}.

%%%%%%%%%%%%%%%%%%%%%%%%%%%%%%%%%%%%%%%%%%%%%%%%%%%%%%%%%%%%%%%%%%%%
\subsection{Active galactic nuclei}
\label{sec:active-galact-nucl}

% ----------------------------------------------------------------
\begin{figure}[t] \centering
  \includegraphics[width=\columnwidth]{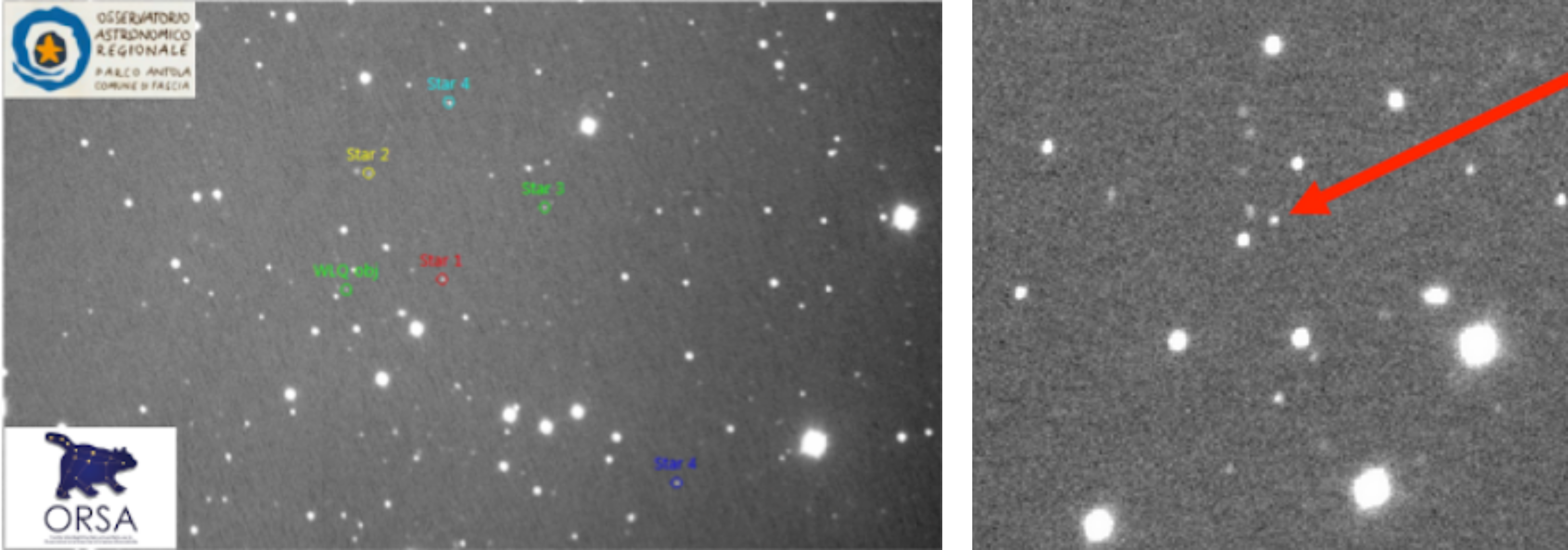}
  \caption{Left: The field of view showing the faint quasar SDSS
    \texttt{J223827.17+135432.6} (labelled WLQ-obj in green) and 5
    reference stars for differential photometry. Right: a zoom around
    the quasar, indicated by a red arrow. }
  \label{fig:qso}
\end{figure}
% ----------------------------------------------------------------

Active galactic nuclei are the most luminous persistent sources of
electromagnetic radiation. Some of them display jets of relativistic
particles and are named blazars when the jet points to the
observer. It is possible to distinguish among various models for
blazars by measuring the optical variability with long monitoring
campaigns~\cite{2014AN....335..417M, 2000AJ....120.1626R,
  2001AJ....122..565R, 2007ApJ...662..182P, 2012MNRAS.420.2899G} . A
feasibility test of such measurements has been performed in order to
make sure that such sources can be observed at \oarpaf. Considering
all the characteristics of the telescope, the instruments, and the
site, several observable blazars have been identified.

A very faint object, \texttt{SDSS J223827.17+135432.6}, with magnitude
$m_R=20$ in the $R$ filter, could be observed (Fig.~\ref{fig:qso})
with a large signal-to-noise ratio of around $10$, with just
$600 \second$ of exposure time~\cite{Righi2015}.
Other interesting blazar candidates that can be observed at \oarpaf
include \texttt{4C+41.11}, \texttt{MG1J021114+1051},
\texttt{PMNJ2227+0037}: these are particularly interesting because
they could also be emitters of very high energy neutrinos.  It would
be of particular interest to observe them in coincidence with
observations in other bands, $X$ or $\gamma$, in order to derive a
multiwavelength measurement.

 For a QSO with magnitude 20, we can achieve a precision
  $\approx 1\%$ in magnitude with an exposure time of around
  $600 \,\second$: this was enough for the planned measurements. In
  the same conditions and with the same set up, achieving an
  uncertainty of $0.01$ mag may require an exposure time a factor 100
  longer, which is not sustainable. However, using the \stx,
  integrated with the AO-X system, we can improve the sensitivity:
  this is a possibility may be explored for measurements requiring a
  better precision on the magnitude.  

%%%%%%%%%%%%%%%%%%%%%%%%%%%%%%%%%%%%%%%%%%%%%%%%%%%%%%%%%%%%%%%%%%%%
\subsection{Gravitationally lensed quasars}
\label{sec:gravitational-lenses}

When a galaxy or cluster of galaxies lies between a far away quasar
and the observer, it produces strong gravitational lensing: multiple
images of the source quasar are observed. Since quasars typically
feature variation in luminosity and color\cite{2011A&A...528A..42R,
  2013A&A...551A.104R}, the various multiple images of the source show
the same features in the light curve, with a time delay due to the
different paths the photons travelled due to the presence of the
lensing galaxy or galaxy cluster.

\begin{figure}[t]
  \centering
  \includegraphics[width=0.7\columnwidth]{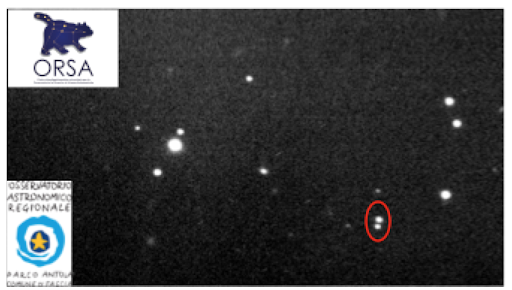}
  \caption{Raw image of the field of view with the
    gravitationally-lensed quasar \texttt{QSO 0957+561}; The red oval
    indicates the two multiple images of the source.  }
  \label{fig:lens}
\end{figure}

Time delays of gravitationally lensed quasars allows measuring the
Hubble parameter~\cite{1964MNRAS.128..295R, 2017MNRAS.468.2590S} :
long campaigns, lasting years, are needed to derive time delays
\cite{2005A&A...436...25E}. The feasibility of such measurements at
\oarpaf has been demonstrated by observing the two lensed quasars SDSS
\texttt{J1004+4112} and \texttt{QSO 0957+561}~\cite{Nicolosi2019}, the
latter shown in Fig.~\ref{fig:lens}. The first is particularly
suitable because it can be observed during most of the year in
Northern Italy, it has a relatively large magnitude, around 18 in the
$I$ band, and it features a large angular separation between the
multiple images, $10$--$30\arcsecond$, permitting to easily separate
and reconstruct the light curves of each image.

A novel experimental method to enhance the number of usable lensed
quasars\footnote{A. Domi,\textit{ et al.}, ``A novel method to measure
  time delay of not resolved gravitationally-lensed quasars using
  source color variations'', in preparation.} for time delay
measurements by $1\meter$ class telescope has been
elaborated~\cite{Nicolosi2019}.  A collaboration with theorists of the
University of Genoa has been established aimed at studying possible
improvements to the relation between time delay and the Hubble
parameter from a theoretical and phenomenological point of
view~\cite{Alchera2017, Alchera2018}.

%%%%%%%%%%%%%%%%%%%%%%%%%%%%%%%%%%%%%%%%%%%%%%%%%%%%%%%%%%%%%%%%%%%
\subsection{Microlensing}
\label{sec:microlensing}

\oarpaf is potentially suited for microlensing studies. In particular,
microlensing searches can be divided into two main categories with
different final goals, briefly explained hereafter.
\begin{itemize}
\item Planetary microlensing: the crowded the stellar field, the
  higher the probability to detect a microlensing event. The best sky
  regions for this activity are the galactic bulge and the Magellanic
  clouds, which are not accessible at \oarpaf.  Although, a derived
  technique called pixel lensing\cite{2010GReGr..42.2101C} can be
  applied in unresolved star regions such as the Andromeda galaxy. The
  use of the AO-X module with the \stx to stabilize the image can be
  promising in this kind of research topic.

\item Quasar microlensing: strongly lensed quasars, described in
  Sect.~\ref{sec:gravitational-lenses}, are affected by the
  microlensing phenomenon. Specifically, massive objects in the
  lensing galaxy, such as stars, impact the observed multiple images
  of the quasar source in the form of sharp and uncorrelated
  brightness variations. These brightness changes are associated with
  the light coming from the innermost region of the quasar, passing
  through a pattern of caustics produced by massive objects in the
  lensing galaxy. It has been demonstrated \cite{microlensingReview}
  that microlensing provides a unique and direct observation of the
  internal structure of the lensed quasar. Such measurement relies on
  the temporal variation of high-magnification caustic crossings which
  vary on timescales of days to years. Moreover, multiwavelength
  observations provide information from distinct emission regions in
  the quasar. Therefore, the monitoring of strongly lensed quasars
  also in terms of microlensing represents a unique and comprehensive
  probe of active black hole structure and dynamics. This method
  requires resolved multiple images of the quasar. Therefore, even
  though this condition reduces the observational sample of the
  \oarpaf telescope, it is still a possible measurement with resolved
  systems such as the ones already observed and cited in
  Sect.~\ref{sec:gravitational-lenses}.
\end{itemize}

%%%%%%%%%%%%%%%%%%%%%%%%%%%%%%%%%%%%%%%%%%%%%%%%%%%%%%%%%%%%%%%%%%%%
\subsection{Asteroids}
\label{sec:asteroids}

% ----------------------------------------------------------------
% ----------------------------------------------------------------

The study of objects in the asteroid belt is an interesting branch of
astronomy due to the relative vicinity of such targets and the
uncertainty on several of their properties and orbital parameters
\footnote{\url{https://nssdc.gsfc.nasa.gov/planetary/planets/asteroidpage.html}}.
It is particularly interesting to measure light curves of the
asteroids when they are in opposition, as this is the most favored
configuration. Simplifying the object as an ellipsoid with 3 axes, one
assumes that luminosity variations are only due to the orientation of
the rotation pole with respect to the ecliptic and the shape of the
ellipsoid. With good quality photometric observations, one can measure
both the orientation of the rotation axis and the ratio between the
major semiaxes of the ellipsoid~\cite{Pospieszalska1985}. In
particular, we plan to observe the asteroid \texttt{1671 Chaika}
\cite{2011ApJ...741...90M}: it is particularly suitable for \oarpaf
because of its large apparent luminosity and interesting orbital
features.

%%%%%%%%%%%%%%%%%%%%%%%%%%%%%%%%%%%%%%%%%%%%%%%%%%%%%%%%%%%%%%%%%%%%
\section{Teaching and outreach}
\label{sec:outreach}

Since the beginning, events for schools and the general public have
been organized at \oarpaf, and had a big success. The astrophile
association Urania is in charge of events for primary school kids and
the citizens, while ORSA manages events for high-school and University
students. So far, only events requiring the physical presence at the
Observatory have been organized because the telescope can only be
operated in local mode.

Data taken with the telescope have been used for teaching purposes in
events for high school and third age students and pictures taken with
the telescope have been shown in various events and festivals. Several
students of the faculty of physics made use of the telescope and its
instruments to perform their master degree theses, actually
greatly contributing to the commissioning and the verification of the
scientific potential~\cite{Righi2015, Cabona2016, Nicolosi2019}. We
expect that, when the facility will be operated remotely, the
  number of events and the number of participants will significantly
  increase and it will also involve potential users of other time
  zones.

%%%%%%%%%%%%%%%%%%%%%%%%%%%%%%%%%%%%%%%%%%%%%%%%%%%%%%%%%%%%%%%%%%%%
\section{Conclusions}
\label{sec:conclusions}

We presented the \oarpaf observatory and its instrumentation.  With
the current setup, we measure a plate scale of
$0.29\arcsecond / \pixel$, a pointing accuracy is of $<10\arcsecond$
rms, and a tracking accuracy of $<1\arcsecond$.
\\
We find for the three available detectors a good linearity range
(approximately from 4\,000 to above 60\,000), and we give an
estimation of the gain and the dark current.
The typical brightness at \oarpaf is found to be
$22.40 m_{AB}/ \arcsecond^2$ in the $B$ filter, down to
$21.14 m_{AB}/ \arcsecond^2$ in the $I$ filter, while the seeing
spans between $1.5$--$3.0\arcsecond$ with a typical value of
$2.5\arcsecond$.
Extinction coefficient and zero points are also calculated by the
observation of standard stars.

\oarpaf instrumentation also includes an échelle and a long slit
spectrograph.  We find for the échelle spectrograph a order
stability of $0.46\pixel$, at 95\% confidence level over one hour, a
dispersion $n$ of the 31 orders between $4$--$105\pixel$ with a width
of $\approx 30\pixel$ and
$n [\nano\meter/\pixel] = 1.39\times 10^{-6}\lambda + 1.45\times 10^{-4}$.
%
% We also find a limit for radial velocity observations of
% $15.8 \kilo\meter / \second$.
%
The commissioning of the long slit spectrograph is in progress.

We foresee that the implementation of the remotization process, of the
instrumentation setup, and of the scientific operations will give a
valuable contribution in cutting-edge scientific topics, such as the
search for exoplanets, the observation of AGNs (these two already
leading to peer-reviewed publications), the measurement of
gravitationally lensed quasars time delays and the study of asteroids.

Thanks to its great potential, the funding needed for the complete
remotization of the facility has been obtained. Therefore, \oarpaf is
expected to fully operate remotely by end of 2021.

%%%%%%%%%%%%%%%%%%%%%%%%%%%%%%%%%%%%%%%%%%%%%%%%%%%%%%%%%%%%%%%%%%%%
\section*{Acknowledgments}
\label{sec:acknowledgments}

We thank the University of Genova for the financial, administrative
and logistic support, in particular dr. W. Riva of the central
administration, as well as all members of ORSA and the students of
DIFI for their enthusiasm. We thank R. Cereseto, M. Cresta and E. Vigo
of the mechanical and electronic services of DIFI and INFN-Sezione di
Genova for the technical interventions on the engines of the eye of
the original dome. We thank the theorists N. Alchera, M. Bonici,
N. Maggiore and L. Panizzi for useful discussions and ideas, as well
as C. Ayala-Loera and S. Brown-Sevilla for the collaboration with
measurements of exoplanetary transits.  We are grateful to L. Nicastro
and E. Palazzi from INAF-OAS for their help in the beginning of the
operations.  We thank the Astelco Systems company for the always
helpful feedback. We are deeply appreciative to Associazione Urania
who made it possible to transform the dream of an observatory on the
Ligurian Apennines in reality. And, of course, we deeply thank Regione
Liguria, Comune di Fascia and Ente Parco Antola for the always
fruitful collaboration.  Fundings for the facility and instruments
were provided by Regione Liguria, Programma Italia-Francia Marittimo,
Comune di Fascia, Ente Parco Antola, Università di Genova, DIFI and
DIBRIS,MIUR Progetto Dipartimenti di Eccellenza. Instruments
for outreach events and activities for students received contributions
by Piano Lauree Scientifiche (PLS) of MIUR. Individuals have received
support by INAF, INFN and MIUR (FFABR). Finally, we thank the
  editor and the anonymous referees, whose remarks contributed to
  improve the paper.

%%%%%%%%%%%%%%%%%%%%%%%%%%%%%%%%%%%%%%%%%%%%%%%%%%%%%%%%%%%%%%%%%%%%
\bibliography{biblio.bib}   % bibliography data in report.bib
\bibliographystyle{spiejour}   % makes bibtex use spiejour.bst

%\end{spacing}

\listoftables
\listoffigures

\end{document}